\documentclass[aps,prl,twocolumn,showpacs,preprintnumbers,amsmath,amssymb]{revtex4}

\usepackage{graphicx}
\usepackage{dcolumn}
\usepackage{bm}

\newcommand{\nc}{\newcommand}
\nc{\be}{\begin{equation}}
\nc{\ee}{\end{equation}}
\nc{\bea}{\begin{eqnarray}}
\nc{\eea}{\end{eqnarray}}
\nc{\bean}{\begin{eqnarray*}}
\nc{\eean}{\end{eqnarray*}}
\nc{\mb}{\mbox}
\nc{\rnc}{\renewcommand}
\nc{\x}{\mb{\boldmath$x$}}
\nc{\A}{\mb{\boldmath$A$}}
\nc{\sa}{\mb{\boldmath$a$}}
\nc{\sss}{\mb{\boldmath$\sigma$}}
\nc{\nab}{\nabla}
\nc{\X}{\sf x}

\begin{document}


\title{Gap evolution in $\nu=1/2$ bilayer quantum Hall systems}
\author{Kentaro Nomura}
 \email{nomura@toki.c.u-tokyo.ac.jp}
\author{Daijiro Yoshioka}%
\affiliation{%
Department of Basic Science, University of Tokyo, 3-8-1 Komaba, Tokyo, Japan
}%


\date{\today}

\begin{abstract}
Fractional quantum Hall states in bilayer system at total filling
 fraction $\nu=1/2$ are examined numerically under some ranges of the
 layer separation and interlayer tunneling. It is shown that the ground
 state changes continuously from two-component state to one-component
 state as the interlayer tunneling rate is increased, while the lowest excited
 state changes discontinuously. This fact explains observed unusual
 behavior of the activation energy which reveals upward cusp as a
 function of interlayer tunneling. Some trial wave functions for the
 ground state and quasihole states are inspected.

\end{abstract}

\maketitle

The fractional quantum Hall effect (FQHE)\cite{tsui} occurs at filling
factor $\nu\equiv n\phi_0^{}/B =p/(2p\pm 1)$ ($p$ integer) in a
two-dimensional electron system with a perpendicular magnetic field $B$,
while the $\nu=1/2$ effect has never been observed\cite{jian}. Here
$\phi_0\equiv hc/e$ is the flux quantum and $n$ is the electron density.
In the physics of the FQHE system, the composite fermion picture is quit
useful\cite{jain}. In this description, the $\nu=p/(2p\pm 1)$ FQHE state
can be understood as the $\nu'=p$ integer quantum Hall effect of
composite fermions which possess two flux quanta $2\phi_0$\cite{jain}.
Thus, $\nu=1/2$ corresponds to zero field system of composite
fermions and the ground state can be understood as the Fermi liquid state of
composite fermions\cite{hlr}.

On the other hand at $\nu=5/2$, which is half filling of the second Landau level, the fractionally quantized plateau of the Hall resistance was observed\cite{will2}.
After some numerical investigations\cite{morf,hr}, the ground state at $\nu=5/2$ is believed to be a kind of $p$-wave paired state of composite fermions first discussed by Moore and Read:\cite{mr}
\bea
\Psi_{\rm MR}={\rm Pf}\left(\frac{1}{u_iv_j-v_iu_j}\right)\prod_{i<j}(u_iv_j-v_iu_j)^2.
\label{moore-read}
\eea
Here we use spherical geometry\cite{sphere} for convenience, $(u_i,v_j)=(\cos(\theta_i/2){\rm e}^{i\phi_i/2},\sin(\theta_i/2){\rm e}^{-i\phi_i/2})$ is spinor coordinate of the $i$'th electron, and
${\rm Pf}[A]$ is the Pfaffian of an antisymmetric matrix $A$. 

 When the internal degree of freedom such as spin or layer index is introduced, the physics becomes more colorful. 
The $\nu=1/2$ FQHE was observed in a double-quantum-well (DQW) structure\cite{eis1} and a wide-single-quantum-well (WSQW)\cite{suen1}.
In a two-layer system without interlayer transfer, the ground state can be approximated as a two-component state proposed by Halperin:\cite{hal}
\bea
 &&\Psi_{331}
\nonumber \\ &&
\ =\prod_{i<j}
(u_iv_j-v_iu_j)^3
\prod_{i<j}(\eta_i\xi_j-\xi_i\eta_j)^{3}  
\prod_{i,j}(u_i\xi_j-v_i\eta_j)^1, \nonumber \\ 
\eea
\noindent
at certain ranges of the ratio of the layer separation and the magnetic length $d/l$\cite{ymg}.
Here $(u_i,v_i)$
and $(\eta_i,\xi_i)$ are complex spinor coordinates of electrons in the top and bottom layer, respectively. 
The experimental result of $d/l$-dependence of the 1/2 FQHE in DQW\cite{eis1} fits the theoretical prediction very well\cite{ymg}.
On the other hand, the 1/2 state measured in WSQW is more
subtle\cite{suen1}, since such a system possesses the duality of a
bilayer and thick single-layer system. 
In fact, both one-component\cite{gww2} and two-component\cite{he}
theoretical models have been proposed.
To see the nature of the ground state in WSQW experimentally, Suen et al. measured activation energy as a function of interlayer tunneling\cite{suen2}.
At $\nu=2/3$ the gap shows downward cusp behavior which indicates a clear transition from two-component to one-component, while at $\nu=1/2$ such a characteristic was not observed.
Since the gap first increases when $\Delta_{SAS}$ is decreased, they considered the two-component state, most likely $\Psi_{331}$-like state, as the ground state.  They also concluded that the two-component state does not evolve to the one-component FQHE state but to a metallic state as $\Delta_{SAS}$ increases.
At the center of the FQHE region, the gap has a sharp upward cusp.

There are also some theoretical progresses.
Based on the pairing picture of composite fermions, Halperin considered the continuous evolution of the ground state between $\Psi_{331}$ and $\Psi_{\rm {MR}}$, and proposed a $d$-$\Delta_{SAS}$ phase diagram\cite{hal2}.
Pursuing this scenario, Ho argued interesting connections between these 1/2 FQHE states and superfluid $^3$He\cite{ho}. Namely $\Psi_{331}$ and $\Psi_{\rm {MR}}$ correspond to the ABM state and A$_1$ state, respectively, and the introduction of $\Delta_{SAS}$ corresponds to the Zeeman splitting along the $x$-axis in $^3$He superfluid. Here the pseudospin $\uparrow$ and $\downarrow$ assign electron in the top layer and bottom layer, respectively.
Such a mean-field picture of composite fermion pairing possesses a beautiful structure. However, the relation to the experimental result mentioned above has been unclear and it stays only in theoretical curiosity.
%
%
In this article we perform a numerical investigation of the evolution of
the $\nu=1/2$ FQHE state as a function of interlayer tunneling and the
layer spacing. It is shown that the ground state evolves continuously,
while the quasihole state changes discontinuously between two-component
and one-component states. Based on these results, a reasonable explanation for inscrutable upward cusp behavior of the activation energy is presented.


\begin{figure}[!b]
\begin{center}
\includegraphics[width=0.47\textwidth]{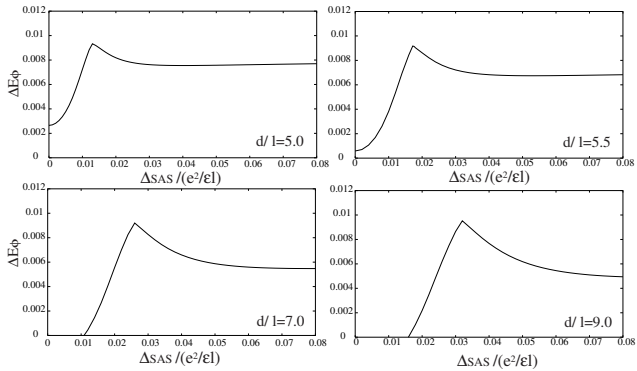}
\caption{
Calculated $\Delta E_{\phi}$ in units of $e^2/\epsilon l$ as a function of
 $\Delta_{SAS}/(e^2/\epsilon l)$ for $d/l=5.0$, $d/l=5.5$, $d/l=7.0$,
 and $d/l=9.0$. The data show upward cusp which is similar to the observed energy gap experimentally. The number of electron is $N=6$.
}
\label{}
\end{center}
\end{figure}

The Hamiltonian is given by
\bea
 H=H_{SAS}+H_{C}.
\eea
The single-particle Hamiltonian;
$H_{SAS}=-\frac{1}{2}\Delta_{SAS}\sum_m(c_{m\uparrow}^{\dag}c_{m\downarrow}^{}+
c_{m\downarrow}^{\dag}c_{m\uparrow}^{}    )= -\Delta_{SAS}S_x$ describes
electron transfer between the layers.
Here the total pseudospin operator is defined as 
${\bf S}=\sum_m
 c_{m\sigma}^{\dag}\frac{{\sss}_{\sigma\sigma'}}{2}c_{m\sigma'}^{}$.
On the other hand $H_{C}$ represents the usual Coulomb interaction within and between the layers:
$H_C=\frac{1}{2}\sum_{m_1\sim m4}\langle m_1,m_2|V_{\sigma\sigma'}|m_3,m_4\rangle c_{m_1\sigma}^{\dag} c_{m_2\sigma'}^{\dag} c_{m_3\sigma'}^{}c_{m_4\sigma}^{}$. 
To model a bilayer system, we consider not only the layer spacing $d$
 but also the thickness $w$. They are treated by using the following form of Coulomb interaction: 
$V_{\uparrow\uparrow}=V_{\downarrow\downarrow}=e^2/\epsilon\sqrt{r^2+w^2}$ and
$V_{\uparrow\downarrow}=V_{\downarrow\uparrow}=e^2/\epsilon\sqrt{r^2+d^2}$.
Since we are interested in changing $d$, the thickness is fixed at $w=3.8l$ in this article.
We first consider the quantity given as
$
 \Delta E_{\phi}=E(N_{\phi}+1,N)+E(N_{\phi}-1,N)-2E(N_{\phi},N) 
$
in $N=6$ electron system as a function of interlayer tunneling where the number of flux $N_{\phi}=2N-3=9$ for the ground state with $p$-wave paired state\cite{ymg,gww}.
As shown in Fig.1 the upward cusp behavior is found at $d/l=5.0,\ 5.5,\ 7.0$, and 9.0. 
The charged excitations are quasiholes or vortices, which effectively contain a half quantum of flux because of pairing.
 Thus to obtain the activation energy, one should divide the value $\Delta E_{\phi}$ by 2 after subtracting the interaction energy for two $e^*=\pm e/4$ charges.
In consideration of this fact, upward cusp behavior of the activation energy observed experimentally is reproduced in our simple theoretical model.

\begin{figure}[!b]
\begin{center}
\includegraphics[width=0.45\textwidth]{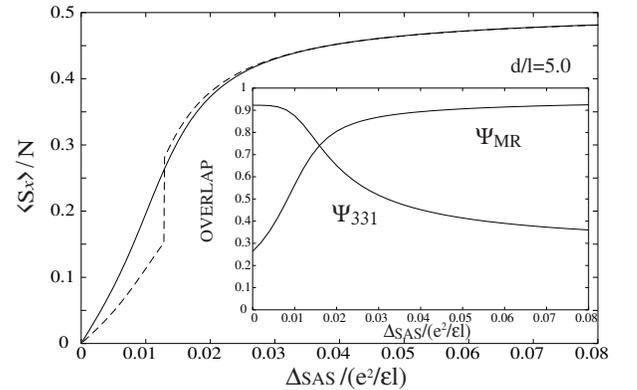}
\caption{ Expectation value of the x-component of the pseudospin in the ground state (solid line) and in the state with an extra flux quantum (dashed line) as a function of $\Delta_{SAS}/(e^2/\epsilon l)$.
Inset: Overlap between the ground state wave function and the Pfaffian, and 331 state trial function as a function of interlayer tunneling in the six-electron system at $d/l=5.0$.  
}
\label{ov}
\end{center}
\end{figure}

\begin{figure}[!b]
\begin{center}
\includegraphics[width=0.45\textwidth]{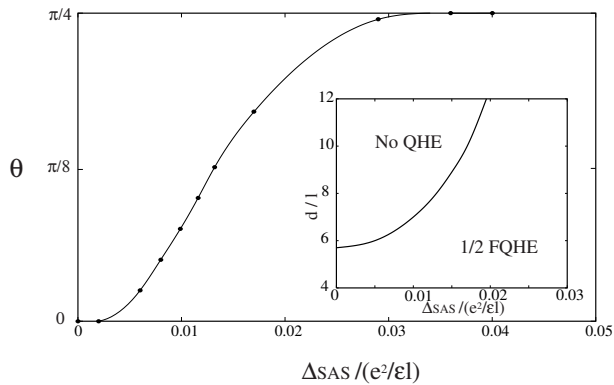}
\caption{Optimized value of $\theta$ which has largest overlap
 $\langle\Psi_{\rm H}[{\bf d}(\theta)]|\Psi_{\rm exact}\rangle$ under the change of $\Delta_{SAS}/(e^2/\epsilon l)$ at $d/l=5.0$. Inset: Calculated quantitative phase diagram of $\nu=1/2$ bilayer system. At the boundary the energy gap collapses.
}
\label{theta}
\end{center}
\end{figure}

Our statement is following. The ground state changes continuously contrary to the case of $\nu=2/3$, while the quasihole state changes discontinuously as a function of $\Delta_{SAS}$. The calculated expectation value $\langle S_x\rangle$ indicates a continuous evolution from two-component ($\langle S_x\rangle=0$) to one-component ($\langle S_x\rangle=N/2$) as shown in Fig\ref{ov}.
Now we show our calculated overlaps between the exact ground state and the trial states as a function of $\Delta_{SAS}$ at $d/l=5.0$ in Fig.\ref{ov}.
The data shows $\Psi_{331}$ and $\Psi_{\rm MR}$ are relevant for small
$\Delta_{SAS}$ and large $\Delta_{SAS}$, respectively.
We found that the crossing point of these two quantities corresponds to the point of upward cusp in the quantity $\Delta E_{\phi}$ at each value of $d/l$.
Thus contrary to the conclusion of ref.\cite{he} or ref.\cite{suen2}, we found that the
upward cusp is related to the crossover between the single component and
two component states.
 Next we inspect the ${\bf d}$-vector description of Ho which is given by\cite{ho}
$
 {\bf d}(\theta)=(0,-i\sin\theta,\cos\theta)
$
 at the middle range of tunneling.
The $p$-wave paired state can be written as
\bea
 \Psi_{\rm H}^{}[{\bf d}(\theta)]={\rm Pf}\left(\frac{\chi[{\bf d}(\theta)]}{u_iv_j-v_iu_j}\right)
\prod_{i<j}(u_iv_j-v_iu_j)^2,
\label{ho}
\eea
where the $2\times 2$ matrix:

\bea
\chi=i[({\bf d}(\theta)\cdot\sss)\sigma_y]=
\left[
\begin{array}{rr}
\cos\theta & \sin\theta \\
\sin\theta & \cos\theta 
\end{array}
\right]
\eea
is related to the order parameter of the triplet pairing\cite{ho,rg}.
Note that $\theta=0$ and $\theta=\pi/4$ correspond $\Psi_{331}$ and $\Psi_{\rm MR}$, respectively.
We constructed the wave function eq.(\ref{ho}) explicitly for several value of $\theta$, and compared with the exact ground state in a six electron system.
The optimized $\theta$ which has largest overlap with the ground state wave function at each value of $\Delta_{SAS}$ is plotted in Fig.\ref{theta}.
The trial state eq.(\ref{ho}) with the optimized $\theta$ has large overlap about 0.9. Note that in triplet pairing the pseudospin is given as ${\bf S}\propto i{\bf d}\times{\bf d}^*$. However the pseudospin operator does not commute with the total Hamiltonian, so above value does not correspond to the exact value calculated in Fig.\ref{ov}.
 It is different from the mean-field theory of composite fermion\cite{ho,rg}.

As we saw in Fig.\ref{ov}, $\langle S_x\rangle$ for the state with one extra flux quantum (dashed
line) has a break in the line at $\Delta_{SAS}/(e^2/\epsilon l )=0.013$,
that indicates a level crossing between two-component and one-component
in the quasihole states.
As we mentioned above, the elementary charged excitations are described as a half quantum vortex $\phi_0^{}/2$ which is called a quasihole.
However, it is impossible to construct numerically the one half flux state. 
So we consider the one extra flux quantum state which should correspond to the two-quasihole state. 
The quasiholes in the Pfaffian state are thought to obey non-abelian statistics\cite{mr}.
For $2n$ quasihole states the $2^{n-1}$ degeneracy is needed to possess non-abelian statistics. Read and Rezayi\cite{rr} confirmed this nature in an exact diagonalization investigation with three body interaction which is the parent Hamiltonian for the Pfaffian state and its quasihole excited states\cite{gww}.
One of the two-quasihole state is the Laughlin type
quasihole\cite{sphere} state: $\prod_{i=1}^{N}v_i\Psi_{\rm MR}$ with the
total angular momentum $L=3$ for a six electron system which corresponds
to the two-quasihole state with zero relative angular momentum. Because of the long-range nature of the Coulomb interaction, such a state with higher total angular momentum should be higher in energy. 
The quantum number of the relative angular momentum of the two-quasiholes must be even because of their statistics. So the state with smaller energy has $L=1$ and the trial function for two-quasiholes is given by\cite{mr,rr}
\begin{figure}[!b]
\includegraphics[width=0.47\textwidth]{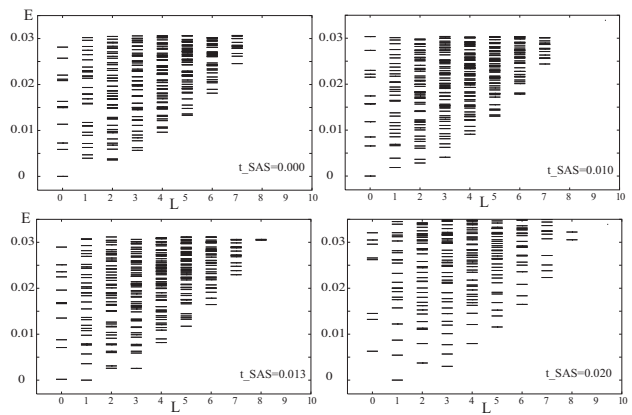}
\caption{Energy eigenvalues [measured in units of $e^2/\epsilon l$] in the six electron system with $N_{\phi}=10$ (extra flux quantum). There is a level crossing when $\Delta_{SAS}/(e^2/\epsilon l)$ increases.}
\label{e}
\end{figure}

\bea
 \Psi_{\rm MR}^{2qh}={\rm Pf}\left(\frac{u_iv_j+v_iu_j}{ u_iv_j-v_iu_j } \right)
\prod_{i<j}(u_iv_j-v_iu_j)^2.
\label{2qhMR}
\eea
One of the quasihole is on the north pole and the other is on the south pole.
We expect such a state becomes relevant in a bilayer system with large $\Delta_{SAS}$. 
Actually $\langle\Psi_{\rm MR}^{2qh}|\Psi\rangle=0.93929$ at
 $\Delta_{SAS}=0.08(e^2/\epsilon l)$ and $d/l=5.0$. 
In the opposite limit of $\Delta_{SAS}\rightarrow 0$ the quasihole states for $\Psi_{331}$ which are given by $\prod_{i=1}^{N_{\uparrow}}v_i\Psi_{331}$ and $\prod_{i=1}^{N_{\downarrow}}\xi_i\Psi_{331}$ might be relevant.
The lowest energy state with two quasiholes should have zero total angular momentum.
Figure \ref{e} shows energy eigenvalues in the six electron system with
 an extra flux as a function of the total angular momentum $L$ and interlayer tunneling $\Delta_{SAS}$.
The state with $L=0$ is lowest in the absence of tunneling. As
$\Delta_{SAS}$ increases energy of the state with $L=1$ decreases and
becomes lowest when $\Delta_{SAS}$ exceeds $0.013(e^2/\epsilon l)$. The latter state can be approximated by $\Psi_{\rm MR}^{2qh}$. The level crossing point exactly corresponds to the cusp point in Fig.1.
In other words, the upward cusp is a sign of the transition between abelian and non-abelian statistics.

\begin{figure}[!t]
\begin{center}
\includegraphics[width=0.41\textwidth]{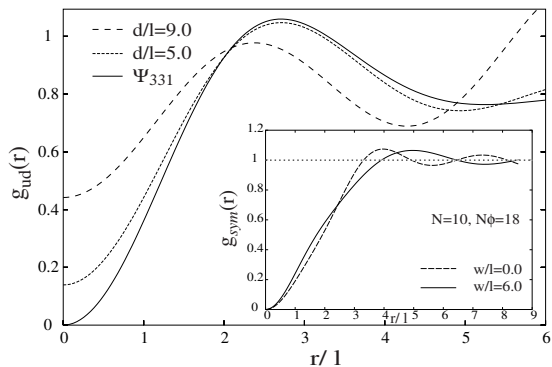}
\caption{ Interlayer pair-correlation function at
 $\Delta_{SAS}=0$, $d/l=5.0$,
 $9.0$ and that in the $\Psi_{331}$ state. The number of electron is $N=6$. Inset: Pair-correlation
 function in the symmetric sector in the case of $d=w= 0$ and
 $d=w=6.0l$ at $\Delta_{SAS}\rightarrow \infty$.
 }
\label{gud}
\end{center}
\end{figure}

\begin{figure}[!t]
\begin{center}
\includegraphics[width=0.45\textwidth]{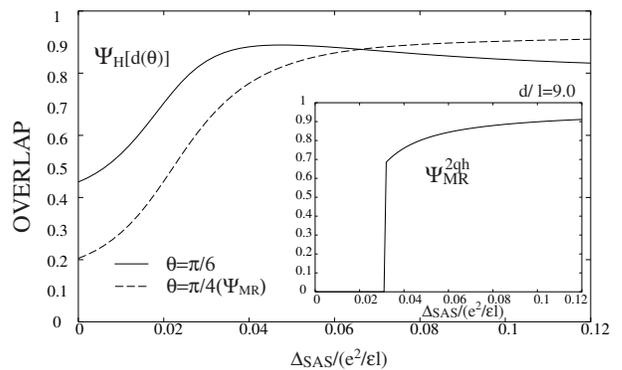}
\caption{Overlap with the trial function eq.(\ref{ho}) with $\theta=\pi/6$ and $\pi/4$ ($\Psi_{\rm MR}$ state) as a function of $\Delta_{SAS}/(e^2/\epsilon l)$ at $d/l=9.0$.
Inset: Overlap with the two-qasihole state of Moore and Read in the added extra quantum flux system. }
\label{ov2}
\end{center}
\end{figure}

Now we study larger $d/l$ region. The $\nu=1/2$ FQHE state is observed in
the WSQW with large $d/l>7$. This fact is far from what was originally expected\cite{ymg}. 
Figure.\ref{gud} shows the interlayer pair-correlation function
$g_{\uparrow\downarrow}(r)$ at $\Delta_{SAS}=0$. In a crude sence, the $\Psi_{331}$ state
can be understood as two Laughlin 1/3 states locked togerther so that
the electrons in one layer are bound to correlation holes in the other.
We can find that correlation hole or locking between the layers declines
and the system goes into  the uncorrelating phase when $d/l$ becomes
large as shown in Fig.6. Thus FQHE state is not realized at $d/l>7$ if $\Delta_{SAS}=0$.
On the other hand, at large $\Delta_{SAS}/(e^2/\epsilon l)$ our
quantitative phase diagram depicted in the inset of Fig.\ref{theta} indicates the 1/2 FQHE.
Actually as shown in Fig.\ref{ov2}, the ground state at large $\Delta_{SAS}$ can be
understood as the Pfaffian state, and the Ho's state (\ref{ho}) with
$\theta=\pi/6$ is relevant around the cusp point. The two-quasihole
state also shows level crossing when the tunneling rate $\Delta_{SAS}$
is increased. 
As Fig.\ref{ov2} shows, at the large $\Delta_{SAS}$ region two-quasihole
state of Moore and Read is relevant.

As a supplement, we comment on the experiment at large $\Delta_{SAS}$.
In the experiment the gap becomes smaller when $\Delta_{SAS}$ is
increased further, and vanishes at $\Delta_{SAS}/(e^2/\epsilon
l)>0.08$\cite{suen2}.
To realize this larger tunneling region, magnetic field is reduced, and
not only $d/l$ but also $w/l$ becoems small ($w/l\sim 2.5$).
 That is to say the well-width $w$ should be appreciated as the third consequent parameter.
When $w$ and $d$ are significantly small and tunneling rate is large,
the system corresponds to a flat single layer system, and the Rezayi-Read state\cite{rezayiread} describing
the Fermi liquid state of composite fermions\cite{hlr} might be relevant,
and quantized plateau does not appear.
In fact, in the limit of $\Delta_{SAS}\rightarrow \infty$ and
$d=w\rightarrow 0$, the two-correlation function in the symmetric sector
(dashed line) shows `$2k_F$-like oscillation'\cite{rezayiread} at $N=10$ and
$N_{\phi}=18$ as we see in Fig.\ref{gud}. Contrary, at $d=w=6.0l$ the short range repulsion is
reduced and the ocsillation disapears that indicates the Cooper
instability. 
Thus to observe the one-component 1/2 FQHE, 
a sample with wide enough well and large tunneling rate should be needed.

In this paper we have investigated the evolution of the $\nu=1/2$
 bilayer FQHE state.
We showed that the ground state evolves continuously as the tunneling rate
is changed, while the quasihole state reveals a level crossing from 
two-component to one-component.
The fact that the energy gap becomes maximum through the transition is
really unusual within the physics of quantum phase transition. 
Perhaps similar singularity might be 
observed in transport phenomena
such as the Hall resistance of drag current or pseudospin current.

This work is supported by a Grant-in-Aid for Scientific Research (C)
10640301 and 14540294 from the Ministry of Education, Science, Sports and Culture.

\end{document}